%% file: main.tex
\documentclass[sigconf]{acmart}
\renewcommand\footnotetextcopyrightpermission[1]{} % removes footnote with conference information in first column
\AtBeginDocument{%
  \providecommand\BibTeX{{%
    \normalfont B\kern-0.5em{\scshape i\kern-0.25em b}\kern-0.8em\TeX}}}

\setcopyright{acmcopyright}
\copyrightyear{2023}
\acmYear{2023}
\acmDOI{XXXXXXX.XXXXXXX}

\acmConference[GLSVLSI'23]{ Proceedings of the Great Lakes Symposium on VLSI}
{June 05--07, 2023}{Knoxville, TN, USA}

\acmPrice{15.00}
\acmISBN{978-1-4503-XXXX-X/18/06}

\usepackage{xspace}
\usepackage{xcolor}
\usepackage[noabbrev,capitalize]{cleveref}
\usepackage[binary-units=true]{siunitx}
\usepackage{threeparttable}
\usepackage{pifont}
\usepackage{fontawesome}
\usepackage{listings}
\usepackage{diagbox}
\usepackage{glossaries}

\newacronym{ecc}{ECC}{error correction code}
\newacronym{pc}{PC}{program counter}
\newacronym{otp}{OTP}{one-time programmable}

\input{results/embench_codesize.tex}

\input{results/embench_runtime.tex}

\newif\ifanonymous
\newcommand{\eg}{e.g.,\ }
\newcommand{\ie}{i.e.,\ }
\newcommand{\cf}{cf.\ }
\newcommand{\otcfi}{SCRAMBLE-CFI\xspace}
\newcommand{\ot}{OpenTitan\xspace}

\newcommand{\asr}{\textbf{AR}\xspace}
\newcommand{\asb}{\textbf{AB}\xspace}
\newcommand{\asm}{\textbf{AM}\xspace}
\newcommand{\asc}{\textbf{AC}\xspace}
\newcommand{\asic}{\textbf{AIC}\xspace}
\newcommand{\cfmf}{\textbf{CFM1}\xspace}
\newcommand{\cfmb}{\textbf{CFM2}\xspace}
\newcommand{\cfma}{\textbf{CFM3}\xspace}
\newcommand{\dtc}{\textbf{DT1}\xspace}
\newcommand{\dtcib}{\textbf{DT1.2a}\xspace}
\newcommand{\dtcr}{\textbf{DT1.2b}\xspace}

\newcommand{\dtcp}{\textbf{DT1.3}\xspace}
\newcommand{\dtcd}{\textbf{DT1.1}\xspace}
\newcommand{\dtnc}{\textbf{DT2}\xspace}
\newcommand{\dti}{\textbf{DT3}\xspace}

\ifanonymous
\newcommand{\repo}{\footnote{Link hidden for blind review.}}
\else
\newcommand{\repo}{\footnote{\url{https://extgit.iaik.tugraz.at/sesys/otcfi}}}
\fi

\definecolor{frenchplum}{RGB}{0,104,181}
\definecolor{frenchred}{RGB}{153,0,0}

\lstdefinelanguage[RISC-V]{Assembler}
{
  alsoletter={.}, % allow dots in keywords
  alsodigit={0x}, % hex numbers are numbers too!
  morekeywords=[1]{ % instructions
    lui, csrrwi, csrrw
  },
  morekeywords=[2]{ % registers
    x0, x28
  },
  morekeywords=[3]{ % registers
    tweak, csr_tweak
  },
  morecomment=[l]{\#},  % as well as # (even though it is unconventional)
}

\definecolor{mauve}{rgb}{0.58,0,0.82}

\begin{document}
\pagestyle{plain} % removes running headers
\title{SCRAMBLE-CFI: Mitigating Fault-Induced Control-Flow\\ Attacks on OpenTitan}

\ifanonymous
\author{Anonymous author(s)}
\else
\author{Pascal Nasahl}
\affiliation{%
  \institution{Graz University of Technology}
  \city{Graz}
  \country{Austria}}
\email{pascal.nasahl@iaik.tugraz.at}

\author{Stefan Mangard}
\affiliation{
  \institution{Graz University of Technology}
  \city{Graz}
  \country{Austria}}
\email{stefan.mangard@iaik.tugraz.at}

\renewcommand{\shortauthors}{Nasahl, et al.}
\fi

\renewcommand{\shorttitle}{SCRAMBLE-CFI: Mitigating Fault-Induced Control-Flow Attacks on OpenTitan}
%------------------------------------------------------------------------------%
\begin{abstract}
%------------------------------------------------------------------------------%
Secure elements physically exposed to adversaries are frequently targeted by fault attacks.
These attacks can be utilized to hijack the control-flow of software allowing the attacker to bypass security measures, extract sensitive data, or gain full code execution.

In this paper, we systematically analyze the threat vector of fault-induced control-flow manipulations on the open-source \ot secure element.
Our thorough analysis reveals that current countermeasures of this chip either induce large area overheads or still cannot prevent the attacker from exploiting the identified threats.
In this context, we introduce \otcfi, an encryption-based control-flow integrity scheme utilizing existing hardware features of \ot.
\otcfi confines, with minimal hardware overhead, the impact of fault-induced control-flow attacks by encrypting each function with a different encryption tweak at load-time.
At runtime, code only can be successfully decrypted when the correct decryption tweak is active. 
We open-source our hardware changes and release our LLVM toolchain automatically protecting programs.
Our analysis shows that \otcfi complementarily enhances security guarantees of \ot with a negligible hardware overhead of less than 3.97\,\% and a runtime overhead of \embenchRUNGEOMEAN for the Embench-IoT benchmarks.
\end{abstract}

\begin{CCSXML}
<ccs2012>
   <concept>
       <concept_id>10002978.10003001.10010777</concept_id>
       <concept_desc>Security and privacy~Hardware attacks and countermeasures</concept_desc>
       <concept_significance>500</concept_significance>
       </concept>
 </ccs2012>
\end{CCSXML}

\ccsdesc[500]{Security and privacy~Hardware attacks and countermeasures}

\keywords{secure element, fault attacks, cryptographic control-flow integrity}

\maketitle

%------------------------------------------------------------------------------%
\section{Introduction}
%------------------------------------------------------------------------------%

In a fault attack, the adversary injects one or multiple bit errors into a chip by using non-invasive methods, such as voltage or clock glitching, or semi-invasive techniques that include shooting with a laser into the silicon of a decapsulated chip~\cite{DBLP:journals/pieee/Bar-ElCNTW06}.
The effects of these bit errors can be exploited and enable the attacker to manipulate the control-flow of software executed on the system~\cite{DBLP:journals/tvlsi/KaraklajicSV13, DBLP:journals/tc/VasselleTMME20, DBLP:conf/fdtc/TimmersSW16, nasahl2019attacking}.

Secure elements, such as \ot~\cite{johnson2018titan}, are used in smartphones and computers as a secure root of trust.
As these chips run security-critical programs, such as a key storage and authentication services, they are lucrative fault targets.
In addition, these devices are easily accessible for fault attackers as they are typically deployed in the wild.
Hence, secure elements need to provide dedicated hardware- and software-based countermeasures protecting the execution of software against faults.

%------------------------------------------------------------------------------%
\subsection*{Contribution}
%------------------------------------------------------------------------------%

In this paper, we first identify threat vectors enabling the adversary to hijack the control-flow of software by inducing faults into \ot.
Subsequently, we thoroughly analyze existing countermeasures aiming to mitigate these attacks.

Our analysis shows that existing countermeasures either induce large area overheads or inadequately reduce the attack surface to address control-flow manipulations.
\otcfi significantly confines the effect of these attacks with a minimal hardware overhead by introducing a cryptographically enforced control-flow integrity scheme. 
In \otcfi, each function is encrypted with a different encryption tweak before the execution of the program.
During runtime, the instrumented program automatically updates the decryption tweak in a CPU register.
Only when the current execution context and the decryption tweak in the register match, the code can be successfully decrypted and executed.
On control-flow redirections to functions outside of the call graph, the tweak mismatches and garbled instructions are decoded, triggering an alert.
As \otcfi utilizes the already existing scrambling unit of \ot for the encryption, our countermeasure only requires minimal hardware changes, yielding an area overhead of less than 3.97\,\%.
Furthermore, our performance analysis shows a small runtime overhead of \embenchRUNGEOMEAN for the Embench-IoT benchmarks.

In summary, our contributions are:
\begin{itemize}
  \item We provide a systematic analysis of threat vectors on \ot allowing an adversary to perform fault-induced control-flow manipulations.
  \item We discuss existing hardware- and software-based fault countermeasures integrated into \ot.
  \item We introduce and open-source\repo \otcfi, an encryption-based control-flow integrity scheme utilizing hardware features of \ot.
  \item Finally, we discuss how \otcfi, complementarily to existing countermeasures, enhances the resilience of \ot against faults with a small runtime and area overhead.
\end{itemize}

%------------------------------------------------------------------------------%
\section{Background}
\label{sec:otcfi:background}
%------------------------------------------------------------------------------%

This section provides background on the \ot chip and control-flow integrity.

%------------------------------------------------------------------------------%
\subsection{\ot}
%------------------------------------------------------------------------------%

\ot is a secure element developed by Google and lowRISC and the entire design, including silicon as well as the firmware, is open-source.
The goal of the project is to develop a chip that acts as a secure root of trust in different computing systems.
\ot contains a rich set of hardware- and software IP, including a key storage and an AES accelerator.

\begin{figure}[t]
  \center
  \includegraphics[width=0.7\linewidth]{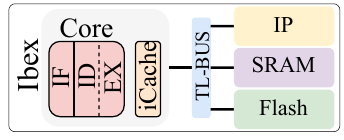}
  \caption{High-level overview of the \ot chip.}
  \label{fig:otcfi:opentitan}
\end{figure}

\Cref{fig:otcfi:opentitan} highlights the main architectural building blocks of the \ot secure element.
The Ibex 32-bit RISC-V processor is connected over the TileLink bus to the program memory (flash), the data memory (SRAM), and several other IP blocks.
To provide data confidentiality, external data stored in flash or SRAM is encrypted by the \ot scrambling unit.
This engine encrypts all data using a round-reduced version of the PRINCE~\cite{DBLP:conf/asiacrypt/BorghoffCGKKKLNPRRTY12} cipher.

%------------------------------------------------------------------------------%
\subsection{Control-Flow Integrity}
%------------------------------------------------------------------------------%

Control-flow integrity schemes aim to detect fault-induced control-flow deviations from the intended control-flow.
Here, these schemes~\cite{DBLP:journals/tr/OhSM02a, DBLP:conf/cgo/ReisCVRA05, DBLP:conf/cosade/SchillingNM22} assign certain points in the program a unique control-flow signature during compile-time.
At runtime, instrumented programs automatically derive the control-flow signature and compare the derived to the predefined signature.
On a mismatch, a control-flow manipulation is detected.
However, as the signature comparison induces performance overhead, the signature checks are only conducted at a coarse granularity, \eg at the function or program end.
Hence, an adversary might still be able to execute security-sensitive code before the control-flow manipulation is detected by CFI.

%------------------------------------------------------------------------------%
\section{Threat Model}
\label{sec:otcfi:threat}
%------------------------------------------------------------------------------%

Our threat model comprises an attacker with physical access to the \ot chip performing fault attacks.
This attacker is capable of injecting single or multiple faults into the secure element by using clock, voltage, or EM glitching techniques or by performing laser fault injection~\cite{DBLP:journals/pieee/Bar-ElCNTW06}.
We assume that these faults cause single or multiple bit-flips~\cite{DBLP:conf/fdtc/VerbauwhedeKS11} in the system.
The goal of the attacker is to redirect the control-flow of software executed on the chip.

%------------------------------------------------------------------------------%
\section{Analysis}
\label{sec:otcfi:analysis}
%------------------------------------------------------------------------------%

In this section, we discuss potential attack vectors within the presumed threat model (\cf \cref{sec:otcfi:threat}) and systematically analyze existing hardware- and software-based countermeasures of \ot aiming to address the identified threats.

%------------------------------------------------------------------------------%
\subsection{Attack Vectors}
\label{sec:otcfi:attackvectors}
%------------------------------------------------------------------------------%

The control-flow of software can be redirected at different control-flow manipulation~(CFM) granularities:
\newline\cfmf:
The attacker redirects the control-flow to a function that cannot be reached from the current execution context, \ie escaping the call graph of the program.
\newline\cfmb:
The control-flow is redirected from one branch target to the other when manipulating conditional branches.
\newline\cfma:
The attacker arbitrarily hijacks the control-flow within a function, \ie to any basic-block.

The attack surface~(A) comprises all elements of \ot (\cf \cref{fig:otcfi:opentitan}).
More specifically, a fault can be injected into CPU internal registers~(\asr), the instruction cache~(\asic), the bus infrastructure~(\asb), or the memory~(\asm).
Furthermore, the attacker can also target the core~(\asc), \ie the instruction fetch, decode, and execute pipeline stages.
For example, when targeting \asc, the attacker can change the behavior of executed instructions when inducing bit flips into the instruction decoder or influence the comparison for a conditional branch in the ALU. 

To manipulate the control-flow, we define three potential data targets~(DT):
\newline\dtc:
\textit{Control-flow} related data comprises relative~(DT1.1) and absolute~(DT1.2) addresses as well as the program counter~(DT1.3).
To flip bits in relative addresses~(DT1.1) used by unconditional and conditional branches, the attacker can inject faults into the immediate field of instructions stored in the program memory~(\asm) or the instruction cache~(\asic) or transferred by the instruction bus~(\asb)~\cite{DBLP:journals/tvlsi/KaraklajicSV13}.
Indirect calls can be manipulated by targeting absolute addresses~(DT1.2a) stored in registers~(\asr) or the data memory~(\asm) or transferred by the data bus~(\asr).
Moreover, returns can be redirected by flipping bits in return addresses~(DT1.2b).
Finally, the attacker also can directly inject faults into the program counter~(DT1.3)(\asr).
Targeting \dtc enables the adversary to arbitrarily manipulate the control-flow, \ie \cfmf, \cfmb, and \cfma.
\newline\dtnc: 
\textit{Non-control-flow} related data, \ie general purpose data, can be faulted by targeting the SRAM~(\asm) or the data bus~(\asb).
When the faulted data is used by conditional branches~\cite{DBLP:journals/tc/VasselleTMME20}, the attacker can influence their execution~(\cfmb).
\newline\dti: \textit{Instructions}, stored in the flash~(\asm) or iCache~(\asic), transferred by the instruction bus~(\asb), and processed by the core~(\asc), can be manipulated by inducing bit flips into the opcode or the operands~\cite{DBLP:conf/fdtc/TimmersSW16, nasahl2019attacking}.
Here, one possible attack would be to skip an instruction by flipping the opcode from a jump~(\texttt{jalr}) to a \texttt{nop}.
Similarly, by manipulating the operand, \eg flipping \texttt{jalr 0(x5)} to \texttt{jalr 0(x6)}, the control-flow can also be arbitrarily redirected.

%------------------------------------------------------------------------------%
\subsection{\ot Countermeasures}
\label{sec:otcfi:countermeasures}
%------------------------------------------------------------------------------%

\paragraph{Hardware-based Countermeasures}

\ot protects security-critical data throughout most of its life cycle using an \gls{ecc}.
More specifically, the integrity of data is protected in the data memory~(\asm), the data bus~(\asb), as well as in the register file~(\asr) of Ibex.
An integrity error allows \ot to detect bit-flips in \dtnc and \dtcib.
Additionally, \gls{ecc} is used in the instruction cache~(\asic), the program memory~(\asm), as well as the instruction bus~(\asb) to detect fault-induced bit-flips, protecting \dti and \dtcd.
To prevent manipulations of the program counter~(\dtcp), Ibex recalculates the derived program counter and triggers an error on a mismatch.

Finally, \ot also provides the possibility of instantiating the Ibex core twice in a lockstep mode.
Here, the execution of the second core is delayed by some cycles and the outputs of the core, \ie the data and instruction interface outputs, are compared.
Although this strategy provides strong protection against faults induced into the pipeline~(\asc), it also more than doubles the area of the CPU. 

\paragraph{Software-based Countermeasures}

In addition to the hardware-based countermeasures, the \ot project also provides software-based fault protection mechanisms that are integrated into a modified LLVM toolchain.
Programs compiled with this toolchain automatically store the return address into a shadow stack.
In the function epilogue, before the return, the current return address is compared to the return address stored in the shadow stack, mitigating bit-flips in return addresses~(\dtcr).
Furthermore, after each indirect branch and return, the compiler inserts an illegal instruction that triggers an exception.
This strategy hinders the adversary from redirecting the control-flow by skipping these instructions~(\dtcp, \dti, and \asc).

%------------------------------------------------------------------------------%
\section{\otcfi}
\label{sec:otcfi:design}
%------------------------------------------------------------------------------%

As shown in the previous section, current \ot countermeasures either induce large area overheads, \ie the dual core lockstep approach, or only provide limited protection against fault-induced control-flow manipulations when targeting the core (\asc).
To that end, in this section, we introduce \otcfi, our control-flow integrity scheme that enhances the resilience of \ot against these attacks with a minimal area and runtime overhead.
Afterwards, in \Cref{sec:otcfi:security}, we then highlight security guarantees and compare \otcfi to existing countermeasures.

%------------------------------------------------------------------------------%
\subsection{Overview}
%------------------------------------------------------------------------------%

\begin{figure}[t]
  \center
  \includegraphics[width=0.85\linewidth]{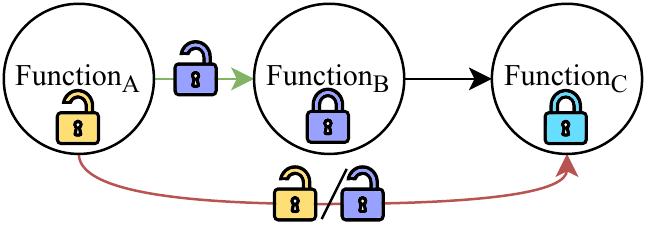}
  \caption{Encrypted call graph.}
  \label{fig:otcfi:enccallgraph}
\end{figure}

In \otcfi, each function is assigned an encryption tweak during program compilation.
Before execution, when loading the program into flash, the code blocks are encrypted with the corresponding tweak.
At runtime, before each function call, the decryption tweak for the call target is placed into a CPU register.
As the modified \ot scramble engine incorporates the content of this register into the decryption, the instructions only can be decrypted when the active tweak matches the tweak determined at compile-time.
\Cref{fig:otcfi:enccallgraph} shows the \otcfi's encrypted call graph, \ie a graph comprising all valid transactions from one function to another.
When redirecting the control-flow from the current execution context to another function encrypted with a different tweak, garbled instructions are fetched and the decoding fails with a high probability.
More specifically, as \otcfi assigns each function, for programs without indirect branches, a unique tweak, any cross-function control-flow manipulation fails.
For programs containing indirect branches, \otcfi guarantees that the attacker cannot redirect the control-flow outside of the call graph.
Summarized, the processor only can execute code blocks when the corresponding decryption tweak is active.

%------------------------------------------------------------------------------%
\subsection{Program Instrumentation}
\label{sec:otcfi:toolchain}
%------------------------------------------------------------------------------%

In order to decrypt the code of a called function, the corresponding decryption tweak needs to be loaded into the tweak register before the control-flow edge, \ie direct and indirect branches.
This program instrumentation is done fully automatically in \otcfi by a modified LLVM~\cite{DBLP:conf/cgo/LattnerA04} RISC-V compiler.

Our custom compiler consists of an analysis and instrumentation pass operating in the backend of the toolchain.
This pass first performs a program analysis to construct the call graph of the program to protect.
Here, we scan each function for direct and indirect calls and determine the call target.
While LLVM already provides this information for direct calls, indirect calls require a points-to analysis~\cite{DBLP:conf/host/NasahlSM21} to reveal the set of potential called functions.

After the extraction of the call graph, we assign each function a unique encryption tweak.
Depending on the number of functions, this tweak is either a \SI{5}{\bit} or \SI{20}{\bit} random number.

\begin{figure}[t]
  \center
  \includegraphics[width=0.9\linewidth]{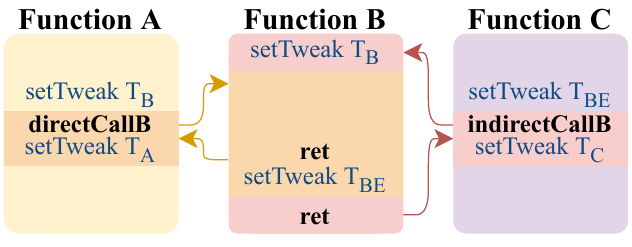}
  \caption{Instrumentation of direct and indirect calls. The different colors highlight code blocks encrypted with different tweaks.}
  \label{fig:otcfi:callinstrumentation}
\end{figure}

\Cref{fig:otcfi:callinstrumentation} shows the instrumentation of direct and indirect calls conducted by the \otcfi compiler.
For the direct call from function \texttt{A} to \texttt{B}, we set the tweak to the tweak of function \texttt{B}, \ie $T_B$.
When returning from this function, the tweak is set back to the tweak used by function \texttt{A}, \ie $T_A$.
Similarly, for indirect branches, \eg from \texttt{C} to \texttt{B}, we also set and reset the tweak before and after the call.
However, instead of using the same tweak as for the direct call, \ie $T_B$, we add an additional entry point to the function, which is encrypted with a different tweak, \ie $T_{BE}$.
In this entry point, we again update the tweak to the tweak for the function to $T_B$.
Moreover, we add a second exit point which is taken when the function was called by an indirect branch.
In this exit point, we set back the tweak to the tweak of the entry point, \ie $T_{BE}$.
The compiler rewrites all addresses used by direct branches to point to the instruction after the added entry point.
Furthermore, in the entry point, we set a flag indicating whether the function returns with the default or the added exit point.

Adding entry points to the program is necessary because indirect calls can have multiple call targets that need to be encrypted with the same tweak.
Without this additional entry point, also direct calls would need to be encrypted with this tweak, allowing the adversary to redirect a direct branch to another function which is called by an indirect branch and is outside of the call graph.

\begin{lstlisting}[language={[RISC-V]Assembler},caption={Tweak update instruction sequence.},label={lst:otcfi:callinstrumentation}]
#5$\,$bit tweak:
 csrrwi  x0, csr_tweak, tweak$_{\textcolor{frenchplum}{5\,\textrm{bit}}}$
#20$\,$bit tweak:
 lui    x28, tweak$_{\textcolor{frenchplum}{20\,\textrm{bit}}}$
 csrrw   x0, csr_tweak, x28
\end{lstlisting}

Listing~\ref{lst:otcfi:callinstrumentation} shows the instruction sequence used to set the \SI{5}{\bit} or \SI{20}{\bit} tweak into the tweak control and status register~(CSR).
For programs with 32 or fewer functions, a single \texttt{csrrwi} instruction loading the \SI{5}{\bit} \texttt{tweak} from the immediate field into the CSR \texttt{csr\_tweak} is sufficient.  

%------------------------------------------------------------------------------%
\subsubsection{Alignment to Encryption Granular}
%------------------------------------------------------------------------------%

As indicated in \cref{fig:otcfi:callinstrumentation}, the next instruction after updating the tweak is already decrypted with this tweak.
Since the decryption granularity of the underlying cipher, \ie \SI{64}{\bit} for PRINCE, does not match the natural RISC-V instruction alignment of 16 or \SI{32}{\bit}, we need to align the set tweak instruction sequence to the decryption granularity.
We conduct this alignment by padding these instructions with \texttt{nops} to \SI{64}{\bit}.

%------------------------------------------------------------------------------%
\subsubsection{Metadata Section}
\label{sec:otcfi:metadata}
%------------------------------------------------------------------------------%

The tweak for each code block is stored in a custom ELF section generated by our toolchain in the \texttt{AsmPrinter} stage.
Here, the compiler emits the start address of each function and the corresponding number of basic-blocks.
Moreover, the offset of each basic-block and the corresponding decryption tweak is emitted into the metadata section by the compiler.
This section is then processed during program deployment.

%------------------------------------------------------------------------------%
\subsection{Program Deployment}
\label{sec:otcfi:flash}
%------------------------------------------------------------------------------%

\begin{figure}[t]
  \center
  \includegraphics[width=\linewidth]{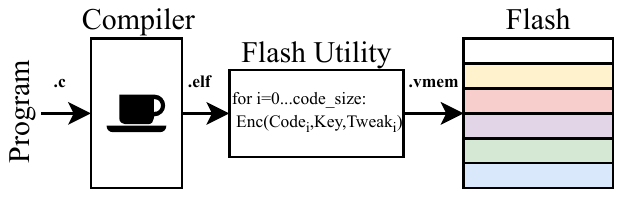}
  \caption{Deployment of protected programs. Code blocks in flash memory encrypted with different \otcfi tweaks are highlighted with different colors.}
  \label{fig:otcfi:flashtool}
\end{figure}

\Cref{fig:otcfi:flashtool} shows the deployment of a program protected with \otcfi on \ot.
Our toolchain first compiles the C source code of the program to an instrumented ELF binary.
Then, the flash utility program converts this binary into a \texttt{VMEM} file, which is loaded into flash memory.

As at the time of writing this paper, the \ot project only provides hardware support for flash scrambling but not the necessary software support and, therefore, disables the scrambling, we extended the flash utility program to encrypt code with the PRINCE cipher.
Here, we encrypt each \SI{64}{\bit} word with the PRINCE cipher using the flash scramble key and the \otcfi tweak.
Note that the flash scramble key is stored inside the \gls{otp} memory.
To retrieve the \otcfi tweak for each code block, the flash utility tool parses the custom metadata section emitted by our toolchain.
Afterwards, the flash is initialized with the encrypted \texttt{VMEM} file and \ot starts the execution of the encrypted code. 

%------------------------------------------------------------------------------%
\subsection{Hardware Changes}
\label{sec:otcfi:hwchanges}
%------------------------------------------------------------------------------%
To realize \otcfi on \ot, minimal-intrusive hardware changes are required:
First~\textit{(i)}, additional control and status registers~(CSRs) need to be added to the Ibex processor.
We implement these registers by using the shadow register primitives provided by the \ot project.
As these shadow registers duplicate the registers and compare the stored values, the content of these CSRs are protected from faults.
By writing to the CSRs, software, \ie binaries compiled with our custom toolchain, can set the current active decryption tweak as well as the lower and upper address bound.
The address range registers comprise the lowest and highest address of program code, which needs to be protected by \otcfi.
Enabling the tweak only for a certain address range is required to access data, such as globals, also stored in the flash memory.
Second~\textit{(ii)}, the tweak needs to be incorporated into the encryption primitive.
One possibility would be to extend the PRINCE cipher to a tweakable block cipher with the \texttt{TWEAKEY}~\cite{DBLP:conf/asiacrypt/JeanNP14} framework.
However, the required \texttt{TWEAKEY} key schedule logic would increase the complexity and area consumption of the cipher.
Moreover, as \otcfi does not need cryptographic strength for control-flow integrity, we inject the tweak into the key using a \texttt{XOR} operation.
This exclusive or is conducted when the address sent to the flash controller is between the lower and upper address stored in the added CSRs.
Otherwise, the tweak is set to $0$ and the PRINCE cipher uses the default key provided by the \gls{otp} controller.
Third~\textit{(iii)}, when the decryption tweak is changed by writing to the CSR, we flush the instruction cache to avoid that scrambled cached instructions are executed.

%------------------------------------------------------------------------------%
\section{Security Analysis \& Comparison}
\label{sec:otcfi:security}
%------------------------------------------------------------------------------%

When the tweak register contains the tweak of the current function, the execution of code fails when the control-flow is redirected to any other function encrypted with a different encryption tweak.
Before a function call, \otcfi updates the tweak register with the tweak of the called function.
Then, the control-flow can only be redirected to this function or functions encrypted with the same tweak.
In \otcfi, the entry points (\cf \cref{sec:otcfi:toolchain}) of functions that an indirect branch can call share the same encryption tweak.
This is inevitable as, at compile-time, the toolchain cannot determine which function gets called by an indirect call at runtime.
However, indirect calls are rare in typical programs and the attacker only can redirect the control-flow to functions that can be reached with this indirect call, \ie are within the call graph.
For programs without indirect calls, \otcfi assigns all functions a unique encryption tweak.
Summarized, for any control-flow redirection outside of the call graph~(\cfmf) the wrong decryption tweak is deterministically used for programs that contain less than $2^{20}$ functions.
For programs that contain more functions than the available tweak space, tweak collisions can occur.

When fetching instructions from program memory with a wrong decryption tweak, garbled instructions are retrieved.
With a high probability, the decoding of these instructions in the instruction decoder pipeline stage fail and an exception is triggered.
However, it could be possible that a garbled instruction again forms a valid instruction.
Nevertheless, \textit{(i)} the probability that the subsequent instructions are also valid is low and \textit{(ii)} usually, the attacker aims to execute a certain instruction in the function and not any that does not trigger an exception.

Note that \otcfi cannot prevent an adversary from manipulating conditional branches~(\cfmb) or from redirecting the control-flow within a function from one basic-block to another~(\cfma).
However, \otcfi could be extended to provide fine-granular protection for highly security-critical code blocks by encrypting code blocks within a function with different encryption keys.

In the current prototype of \otcfi, we incorporate the tweak into the encryption by XORing it to the encryption key.
This XOR creates a dependency of key and tweak, allowing an attacker to potentially learn about the key when the tweak is known.
However, the binary (including the tweaks) is inaccessible to an adversary as it is stored in the encrypted flash.

%------------------------------------------------------------------------------%
\subsection{Security Comparison}
\label{sec:otcfi:seccomparison}
%------------------------------------------------------------------------------%

\Cref{tab:otcfi:mapping} highlights protection guarantees of different hardware- and software-based \ot countermeasures and \otcfi against faults into different attack targets.

\begin{table}[b]
  \small
  \caption{Protection guarantees of different countermeasure when targeting different attack surfaces.}
  \label{tab:otcfi:mapping}
  \begin{threeparttable}
  \begin{tabular}{lccccc}
  \hline
                          & \multicolumn{5}{c}{\textbf{Attack Targets}}                          \\
  \textbf{Countermeasure} & \textbf{AR} & \textbf{AIC} & \textbf{AB} & \textbf{AM} & \textbf{AC} \\ \hline
  Data memory ECC         & -           & -            & -           & \ding{52}   & -           \\
  Program memory ECC      & -           & -            & -           & \ding{52}   & -           \\
  Data bus ECC            & -           & -            & \ding{52}   & -           & -           \\
  Instruction bus ECC     & -           & -            & \ding{52}   & -           & -           \\
  Register file ECC       & \ding{52}   & -            & -           & -           & -           \\
  iCache ECC              & -           & \ding{52}    & -           & -           & -           \\
  PC protection           & \ding{52}   & -            & -           & -           & \ding{52}   \\
  SW-based defense        & \faShield   & -            & -           & -           & \faShield   \\
  Dual core lockstep      & \ding{52}   & \ding{52}    & -           & -           & \ding{52}   \\
  \otcfi                  & \faShield   & \faShield    & \faShield   & \faShield   & \faShield   \\ \hline
\end{tabular}
\small{
\begin{center}
    \ding{52} \quad Full \quad
    \faShield \quad Partial \quad
    - \quad No Protection
\end{center}
}
\end{threeparttable}
\end{table}

The \gls{ecc}-based countermeasures provide full protection when inducing faults into their protection domain.
For example, bit-flips induced into instructions~(\dti) stored in the instruction cache~(\asic) or the program memory~(\asm) can be detected reliably.
As the \gls{pc} protection recalculates and compares the \gls{pc} to the current \gls{pc}, also faults into the core~(\asc) can be detected.
The software-based defense integrated into the custom LLVM toolchain only provides partial protection, \ie only control-flow manipulations aiming to manipulate return addresses~(\dtcr) or skipping certain instructions~(\dti) can be detected.
As shown in \cref{tab:otcfi:mapping}, from the existing countermeasures, only the lockstep approach can provide strong protection against faults induced into the core~(\asc).
However, this strategy also induces a high area overhead as the Ibex core needs to be instantiated twice and an error detection logic needs to be added.
In comparison, \otcfi provides strong protection against control-flow manipulations for all attack targets, even when targeting the core, with minimal hardware overhead.
Especially when combined with the other existing countermeasures, \otcfi significantly minimizes the attack surface and allows \ot to withstand most of the control-flow attacks.
Combined, \ot is protected against arbitrary control-flow manipulations~(\cfmf, \cfmb, and \cfma) when targeting all attack targets except the core~(\asc).
When injecting faults into \asc, \otcfi fills the protection gap of hindering an adversary from redirecting the control-flow outside of the call graph~(\cfmf).
However, when also \cfmb and \cfma need to be mitigated, \otcfi needs to be deployed on a finer granularity (\cf \cref{sec:otcfi:security}) or the lockstep approach needs to be installed.
Summarized, we argue that \otcfi provides a good area-security tradeoff allowing \ot to withstand a multitude of different fault attacks.

%------------------------------------------------------------------------------%
\section{Performance Overhead}
\label{sec:otcfi:performance}
%------------------------------------------------------------------------------%

To evaluate the performance and code size overhead, we compiled the Embench-IoT~\cite{embenchiot} benchmarks with our custom toolchain.
We excluded benchmarks requiring libraries, such as \texttt{math} and \texttt{string}, currently not provided by the \ot framework.

%\begin{table}[b]
%  \small
%  \centering
%  \caption{Code size overhead for the Embench-IoT benchmarks.}
%  \label{tab:otcfi:embenchcodesize}
%  \begin{tabular}{lc}
%  \hline
%  \textbf{Benchmark}  & \textbf{Overhead {[}\%{]}} \\ \hline
%  aha-mont64          & \embenchCODEahamont        \\
%  crc32               & \embenchCODEcrc            \\
%  edn                 & \embenchCODEedn            \\
%  huffbench           & \embenchCODEhuffbench      \\
%  matmult             & \embenchCODEmatmult        \\ 
%  md5                 & \embenchCODEmd             \\ 
%  nettle-aes          & \embenchCODEnettleaes      \\ 
%  nettle-sha256       & \embenchCODEnettlesha      \\ 
%  nsichneu            & \embenchCODEnsichneu       \\ 
%  picojpeg            & \embenchCODEpicojpeg       \\ 
%  primecount          & \embenchCODEprimecount     \\ 
%  sglib               & \embenchCODEsglib          \\ 
%  tarfind             & \embenchCODEtarfind        \\ \hline
%  \textbf{Geometric mean} & \embenchCODEGEOMEAN    \\ \hline
%  \end{tabular}
%\end{table}

To measure the code size overhead, we compared the protected binaries to the unprotected baseline with the GNU \texttt{size} utility.
%\Cref{tab:otcfi:embenchcodesize} highlights the percentual code size overhead measured with the GNU \texttt{size} utility when compared to the unprotected baseline.
Our analysis shows a code size overhead between \embenchCODEMIN and \embenchCODEMAX and a geometric mean of \embenchCODEGEOMEAN.
This overhead comprises the \textit{(i)} inserted tweak switch instructions, the \textit{(ii)} alignment of these instructions to the encryption granular, and the \textit{(iii)} metadata binary section.
As this section only is needed by the flash utility program to encrypt code blocks with different encryption tweaks (\cf \cref{sec:otcfi:flash}), this metadata is not stored in the flash memory.

\begin{figure*}[h!]
  \center
  \includegraphics[width=0.9\linewidth]{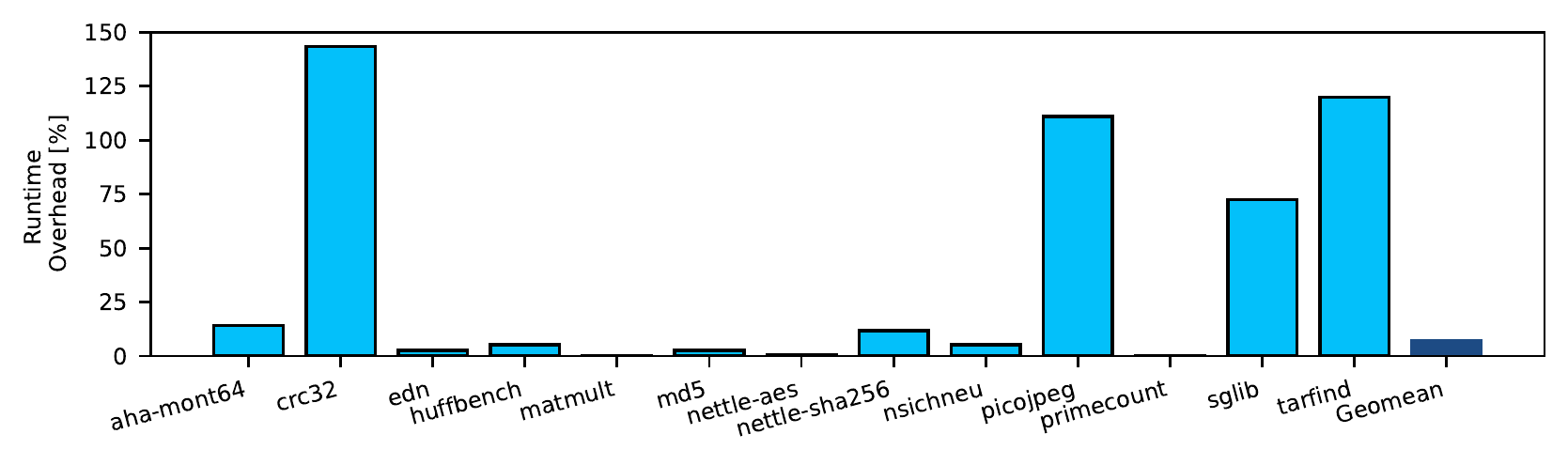}
  \caption{Runtime overhead for the Embench-IoT benchmarks.}
  \label{fig:otcfi:embenchruntime}
\end{figure*}

In order to analyze the performance impact of binaries instrumented with \otcfi, we measured the
CPU cycles by reading the \texttt{mcycle} CSR of Ibex.
Here, we executed the protected and unprotected binaries on a cycle-accurate Verilator model of \ot.
As shown in \cref{fig:otcfi:embenchruntime}, we measured a runtime overhead between \embenchRUNMIN and \embenchRUNMAX and a geometric mean of \embenchRUNGEOMEAN.
Benchmarks, such as \texttt{crc32} or \texttt{tarfind}, frequently calling small functions induce larger runtime overheads than benchmarks only performing a small number of function calls.

%------------------------------------------------------------------------------%
\section{Area Overhead}
\label{sec:otcfi:area}
%------------------------------------------------------------------------------%

The public \ot hardware design flow currently only allows to synthesize the Ibex core with open-source synthesis tools.
Therefore, we synthesized the Ibex processor with the Yosys open synthesis suite and the Nangate \SI{45}{nm} standard cell library to analyze the area overhead introduced by our hardware changes. 
Here, we measured an area increase from 26.48\,kGE to 27.53\,kGE (3.97\,\%).
These hardware changes comprise the additional CSRs as well as the address range comparison.
Note that the XOR of the tweak with the key is conducted in the flash controller and is currently not reflected in the hardware overhead number.
According to the synthesis logs created with the proprietary design flow published by the \ot project~\cite{otbuild}, the Ibex occupies 3.4\,\% of the overall chip area.
Hence, the hardware overhead induced by \otcfi to the overall area is negligible.

%------------------------------------------------------------------------------%
\section{Related Work}
\label{sec:otcfi:relwork}
%------------------------------------------------------------------------------%

Currently, encryption-based control-flow integrity schemes, such as SOFIA~\cite{DBLP:journals/compsec/ClercqGUMV17} or SCFP~\cite{DBLP:conf/eurosp/WernerUSM18}, require intrusive hardware changes in the processor's pipeline to realize their protection.
Hence, when integrating these changes into a chip, an extensive re-verification of the entire processor is needed.
Although EC-CFI~\cite{eccfi} provides cryptographic CFI on Intel hardware without hardware modifications, the measured runtime overheads are high.
In comparison, \otcfi induces small runtime overheads with minimal hardware changes that allow designers to re-verify their design with minimal effort.

%------------------------------------------------------------------------------%
\section{Conclusion}
\label{sec:otcfi:conclusion}
%------------------------------------------------------------------------------%

In this paper, we thoroughly analyzed fault threat vectors allowing an adversary to manipulate the control-flow of software executed on \ot.
We provided an overview of current \ot countermeasures and discussed their protection capabilities.
Furthermore, we introduced \otcfi, which mitigates fault attacks aiming to redirect the control-flow outside of the call graph.
We showcased that \otcfi is a strong security addition to existing countermeasures inducing minimal runtime and area overheads.

%------------------------------------------------------------------------------%
\section{Acknowledgments}
%------------------------------------------------------------------------------%
\ifanonymous
Intentionally left blank.
\else
This project has received funding from the Austrian Research Promotion Agency (FFG) via the AWARE project (grant number 891092).
\fi
\bibliographystyle{ACM-Reference-Format}
\bibliography{bibliography}

\end{document}

%% file: results/embench_codesize.tex
\newcommand{\embenchCODEGEOMEAN}{1.69\,\%\xspace}
\newcommand{\embenchCODEMIN}{0.74\,\%\xspace}
\newcommand{\embenchCODEMAX}{8.88\,\%\xspace}

%% file: results/embench_runtime.tex
\newcommand{\embenchRUNGEOMEAN}{7.02\,\%\xspace}
\newcommand{\embenchRUNMIN}{0.22\,\%\xspace}
\newcommand{\embenchRUNMAX}{143.35\,\%\xspace}

%% file: main.bbl

\begin{thebibliography}{19}

%%% the Web address.

\ifx \showCODEN    \undefined \def \showCODEN     #1{\unskip}     \fi
\ifx \showDOI      \undefined \def \showDOI       #1{#1}\fi
\ifx \showISBNx    \undefined \def \showISBNx     #1{\unskip}     \fi
\ifx \showISBNxiii \undefined \def \showISBNxiii  #1{\unskip}     \fi
\ifx \showISSN     \undefined \def \showISSN      #1{\unskip}     \fi
\ifx \showLCCN     \undefined \def \showLCCN      #1{\unskip}     \fi
\ifx \shownote     \undefined \def \shownote      #1{#1}          \fi
\ifx \showarticletitle \undefined \def \showarticletitle #1{#1}   \fi
\ifx \showURL      \undefined \def \showURL       {\relax}        \fi
% invisible to TeX
\providecommand\bibfield[2]{#2}
\providecommand\bibinfo[2]{#2}
\providecommand\natexlab[1]{#1}
\providecommand\showeprint[2][]{arXiv:#2}

\bibitem[Bar{-}El et~al\mbox{.}(2006)]%
        {DBLP:journals/pieee/Bar-ElCNTW06}
\bibfield{author}{\bibinfo{person}{Hagai Bar{-}El}, \bibinfo{person}{Hamid
  Choukri}, \bibinfo{person}{David Naccache}, \bibinfo{person}{Michael
  Tunstall}, {and} \bibinfo{person}{Claire Whelan}.}
  \bibinfo{year}{2006}\natexlab{}.
\newblock \showarticletitle{{The Sorcerer's Apprentice Guide to Fault
  Attacks}}.
\newblock \bibinfo{journal}{\emph{Proc. {IEEE}}}  \bibinfo{volume}{94}
  (\bibinfo{year}{2006}), \bibinfo{pages}{370--382}.
\newblock


\bibitem[Borghoff et~al\mbox{.}(2012)]%
        {DBLP:conf/asiacrypt/BorghoffCGKKKLNPRRTY12}
\bibfield{author}{\bibinfo{person}{Julia Borghoff}, \bibinfo{person}{Anne
  Canteaut}, \bibinfo{person}{Tim G{\"{u}}neysu}, \bibinfo{person}{Elif~Bilge
  Kavun}, \bibinfo{person}{Miroslav Knezevic}, \bibinfo{person}{Lars~R.
  Knudsen}, \bibinfo{person}{Gregor Leander}, \bibinfo{person}{Ventzislav
  Nikov}, \bibinfo{person}{Christof Paar}, \bibinfo{person}{Christian
  Rechberger}, \bibinfo{person}{Peter Rombouts}, \bibinfo{person}{S{\o}ren~S.
  Thomsen}, {and} \bibinfo{person}{Tolga Yal{\c{c}}in}.}
  \bibinfo{year}{2012}\natexlab{}.
\newblock \showarticletitle{{{PRINCE} - {A} Low-Latency Block Cipher for
  Pervasive Computing Applications - Extended Abstract}}. In
  \bibinfo{booktitle}{\emph{{ASIACRYPT}}} \emph{(\bibinfo{series}{LNCS},
  Vol.~\bibinfo{volume}{7658})}. \bibinfo{pages}{208--225}.
\newblock
\showISBNx{978-3-642-34960-7}


\bibitem[de~Clercq et~al\mbox{.}(2017)]%
        {DBLP:journals/compsec/ClercqGUMV17}
\bibfield{author}{\bibinfo{person}{Ruan de Clercq}, \bibinfo{person}{Johannes
  G{\"{o}}tzfried}, \bibinfo{person}{David {\"{U}}bler},
  \bibinfo{person}{Pieter Maene}, {and} \bibinfo{person}{Ingrid Verbauwhede}.}
  \bibinfo{year}{2017}\natexlab{}.
\newblock \showarticletitle{{{SOFIA:} Software and control flow integrity
  architecture}}.
\newblock \bibinfo{journal}{\emph{Comput. Secur.}}  \bibinfo{volume}{68}
  (\bibinfo{year}{2017}), \bibinfo{pages}{16--35}.
\newblock


\bibitem[Jean et~al\mbox{.}(2014)]%
        {DBLP:conf/asiacrypt/JeanNP14}
\bibfield{author}{\bibinfo{person}{J{\'{e}}r{\'{e}}my Jean},
  \bibinfo{person}{Ivica Nikolic}, {and} \bibinfo{person}{Thomas Peyrin}.}
  \bibinfo{year}{2014}\natexlab{}.
\newblock \showarticletitle{{Tweaks and Keys for Block Ciphers: The {TWEAKEY}
  Framework}}. In \bibinfo{booktitle}{\emph{{ASIACRYPT}}}
  \emph{(\bibinfo{series}{LNCS}, Vol.~\bibinfo{volume}{8874})}.
  \bibinfo{pages}{274--288}.
\newblock
\showISBNx{978-3-662-45607-1}


\bibitem[Johnson et~al\mbox{.}(2018)]%
        {johnson2018titan}
\bibfield{author}{\bibinfo{person}{Scott Johnson}, \bibinfo{person}{Dominic
  Rizzo}, \bibinfo{person}{Parthasarathy Ranganathan}, \bibinfo{person}{Jon
  McCune}, {and} \bibinfo{person}{Richard Ho}.}
  \bibinfo{year}{2018}\natexlab{}.
\newblock \showarticletitle{Titan: enabling a transparent silicon root of trust
  for cloud}. In \bibinfo{booktitle}{\emph{Hot Chips: A Symposium on High
  Performance Chips}}, Vol.~\bibinfo{volume}{194}.
\newblock


\bibitem[Karaklajic et~al\mbox{.}(2013)]%
        {DBLP:journals/tvlsi/KaraklajicSV13}
\bibfield{author}{\bibinfo{person}{Dusko Karaklajic},
  \bibinfo{person}{J{\"{o}}rn{-}Marc Schmidt}, {and} \bibinfo{person}{Ingrid
  Verbauwhede}.} \bibinfo{year}{2013}\natexlab{}.
\newblock \showarticletitle{{Hardware Designer's Guide to Fault Attacks}}.
\newblock \bibinfo{journal}{\emph{{IEEE} Trans. Very Large Scale Integr.
  Syst.}}  \bibinfo{volume}{21} (\bibinfo{year}{2013}),
  \bibinfo{pages}{2295--2306}.
\newblock


\bibitem[Lattner and Adve(2004)]%
        {DBLP:conf/cgo/LattnerA04}
\bibfield{author}{\bibinfo{person}{Chris Lattner} {and}
  \bibinfo{person}{Vikram~S. Adve}.} \bibinfo{year}{2004}\natexlab{}.
\newblock \showarticletitle{{{LLVM:} {A} Compilation Framework for Lifelong
  Program Analysis {\&} Transformation}}. In \bibinfo{booktitle}{\emph{{CGO}}}.
  \bibinfo{pages}{75--88}.
\newblock
\showISBNx{0-7695-2102-9}


\bibitem[Nasahl et~al\mbox{.}(2021)]%
        {DBLP:conf/host/NasahlSM21}
\bibfield{author}{\bibinfo{person}{Pascal Nasahl}, \bibinfo{person}{Robert
  Schilling}, {and} \bibinfo{person}{Stefan Mangard}.}
  \bibinfo{year}{2021}\natexlab{}.
\newblock \showarticletitle{{Protecting Indirect Branches Against Fault Attacks
  Using {ARM} Pointer Authentication}}. In \bibinfo{booktitle}{\emph{{HOST}}}.
  \bibinfo{pages}{68--79}.
\newblock
\showISBNx{978-1-6654-1357-2}


\bibitem[Nasahl et~al\mbox{.}(2023)]%
        {eccfi}
\bibfield{author}{\bibinfo{person}{Pascal Nasahl}, \bibinfo{person}{Salmin
  Sultana}, \bibinfo{person}{Hans Liljestrand}, \bibinfo{person}{Karanvir
  Grewal}, \bibinfo{person}{Michael LeMay}, \bibinfo{person}{David~M. Durham},
  \bibinfo{person}{David Schrammel}, {and} \bibinfo{person}{Stefan Mangard}.}
  \bibinfo{year}{2023}\natexlab{}.
\newblock \showarticletitle{{{EC-CFI:} Control-Flow Integrity via Code
  Encryption Counteracting Fault Attacks}}.
\newblock \bibinfo{journal}{\emph{CoRR}}  \bibinfo{volume}{abs/2301.13760}
  (\bibinfo{year}{2023}).
\newblock


\bibitem[Nasahl and Timmers(2019)]%
        {nasahl2019attacking}
\bibfield{author}{\bibinfo{person}{Pascal Nasahl} {and} \bibinfo{person}{Niek
  Timmers}.} \bibinfo{year}{2019}\natexlab{}.
\newblock \showarticletitle{Attacking {AUTOSAR} using software and hardware
  attacks}.
\newblock \bibinfo{journal}{\emph{escar USA}} (\bibinfo{year}{2019}).
\newblock


\bibitem[Oh et~al\mbox{.}(2002)]%
        {DBLP:journals/tr/OhSM02a}
\bibfield{author}{\bibinfo{person}{Nahmsuk Oh}, \bibinfo{person}{Philip~P.
  Shirvani}, {and} \bibinfo{person}{Edward~J. McCluskey}.}
  \bibinfo{year}{2002}\natexlab{}.
\newblock \showarticletitle{{Control-flow checking by software signatures}}.
\newblock \bibinfo{journal}{\emph{{IEEE} Trans. Reliab.}}  \bibinfo{volume}{51}
  (\bibinfo{year}{2002}), \bibinfo{pages}{111--122}.
\newblock


\bibitem[OpenTitan(2023)]%
        {otbuild}
\bibfield{author}{\bibinfo{person}{OpenTitan}.}
  \bibinfo{year}{2023}\natexlab{}.
\newblock \bibinfo{title}{CHIP\_EARLGREY\_ASIC Synthesis Results}.
\newblock
  \bibinfo{howpublished}{\url{https://reports.opentitan.org/hw/top_earlgrey/syn/2022.07.02_00.42.20/results.html}}.
\newblock


\bibitem[Patterson et~al\mbox{.}(2023)]%
        {embenchiot}
\bibfield{author}{\bibinfo{person}{David Patterson}, \bibinfo{person}{Jeremy
  Bennett}, \bibinfo{person}{Palmer Dabbelt}, \bibinfo{person}{Cesare Garlati},
  \bibinfo{person}{G.~S. Madhusudan}, {and} \bibinfo{person}{Trevor Mudge}.}
  \bibinfo{year}{2023}\natexlab{}.
\newblock \bibinfo{title}{Embench: Open Benchmarks for Embedded Platforms}.
\newblock \bibinfo{howpublished}{\url{https://www.embench.org/}}.
\newblock


\bibitem[Reis et~al\mbox{.}(2005)]%
        {DBLP:conf/cgo/ReisCVRA05}
\bibfield{author}{\bibinfo{person}{George~A. Reis}, \bibinfo{person}{Jonathan
  Chang}, \bibinfo{person}{Neil Vachharajani}, \bibinfo{person}{Ram Rangan},
  {and} \bibinfo{person}{David~I. August}.} \bibinfo{year}{2005}\natexlab{}.
\newblock \showarticletitle{{{SWIFT:} Software Implemented Fault Tolerance}}.
  In \bibinfo{booktitle}{\emph{{CGO}}}. \bibinfo{pages}{243--254}.
\newblock
\showISBNx{0-7695-2298-X}


\bibitem[Schilling et~al\mbox{.}(2022)]%
        {DBLP:conf/cosade/SchillingNM22}
\bibfield{author}{\bibinfo{person}{Robert Schilling}, \bibinfo{person}{Pascal
  Nasahl}, {and} \bibinfo{person}{Stefan Mangard}.}
  \bibinfo{year}{2022}\natexlab{}.
\newblock \showarticletitle{{{FIPAC:} Thwarting Fault- and Software-Induced
  Control-Flow Attacks with {ARM} Pointer Authentication}}. In
  \bibinfo{booktitle}{\emph{{COSADE}}} \emph{(\bibinfo{series}{LNCS},
  Vol.~\bibinfo{volume}{13211})}. \bibinfo{pages}{100--124}.
\newblock
\showISBNx{978-3-030-99765-6}


\bibitem[Timmers et~al\mbox{.}(2016)]%
        {DBLP:conf/fdtc/TimmersSW16}
\bibfield{author}{\bibinfo{person}{Niek Timmers}, \bibinfo{person}{Albert
  Spruyt}, {and} \bibinfo{person}{Marc Witteman}.}
  \bibinfo{year}{2016}\natexlab{}.
\newblock \showarticletitle{{Controlling {PC} on {ARM} Using Fault Injection}}.
  In \bibinfo{booktitle}{\emph{{FDTC}}}. \bibinfo{pages}{25--35}.
\newblock
\showISBNx{978-1-5090-1108-7}


\bibitem[Vasselle et~al\mbox{.}(2020)]%
        {DBLP:journals/tc/VasselleTMME20}
\bibfield{author}{\bibinfo{person}{Aur{\'{e}}lien Vasselle},
  \bibinfo{person}{Hugues Thiebeauld}, \bibinfo{person}{Quentin Maouhoub},
  \bibinfo{person}{Ad{\`{e}}le Morisset}, {and}
  \bibinfo{person}{S{\'{e}}bastien Ermeneux}.} \bibinfo{year}{2020}\natexlab{}.
\newblock \showarticletitle{{Laser-Induced Fault Injection on Smartphone
  Bypassing the Secure Boot-Extended Version}}.
\newblock \bibinfo{journal}{\emph{{IEEE} Trans. Computers}}
  \bibinfo{volume}{69} (\bibinfo{year}{2020}), \bibinfo{pages}{1449--1459}.
\newblock


\bibitem[Verbauwhede et~al\mbox{.}(2011)]%
        {DBLP:conf/fdtc/VerbauwhedeKS11}
\bibfield{author}{\bibinfo{person}{Ingrid Verbauwhede}, \bibinfo{person}{Dusko
  Karaklajic}, {and} \bibinfo{person}{J{\"{o}}rn{-}Marc Schmidt}.}
  \bibinfo{year}{2011}\natexlab{}.
\newblock \showarticletitle{{The Fault Attack Jungle - {A} Classification Model
  to Guide You}}. In \bibinfo{booktitle}{\emph{{FDTC}}}. \bibinfo{pages}{3--8}.
\newblock
\showISBNx{978-1-4577-1463-4}


\bibitem[Werner et~al\mbox{.}(2018)]%
        {DBLP:conf/eurosp/WernerUSM18}
\bibfield{author}{\bibinfo{person}{Mario Werner}, \bibinfo{person}{Thomas
  Unterluggauer}, \bibinfo{person}{David Schaffenrath}, {and}
  \bibinfo{person}{Stefan Mangard}.} \bibinfo{year}{2018}\natexlab{}.
\newblock \showarticletitle{{Sponge-Based Control-Flow Protection for IoT
  Devices}}. In \bibinfo{booktitle}{\emph{{EURO S{\&}P}}}.
  \bibinfo{pages}{214--226}.
\newblock
\showISBNx{978-1-5386-4228-3}


\end{thebibliography}
